\documentclass{emulateapj}
\usepackage{color}

\usepackage{graphicx}
\usepackage[space]{grffile}
\usepackage{latexsym}
\usepackage{amsfonts,amsmath,amssymb}
\usepackage{url}
\usepackage[utf8x]{inputenc}
\usepackage{hyperref}
\hypersetup{colorlinks=false,pdfborder={0 0 0}}
\usepackage{textcomp}
\usepackage{longtable}
\usepackage{multirow,booktabs}

\usepackage{natbib}

\def\simplesize{9cm}
\def\doublesize{18cm}

\shorttitle{Solar Wind Turbulence and Phase Coherence at Ion Kinetic Scales}
\shortauthors{Lion et al.}

\begin{document}

\title{Coherent events and spectral shape at ion kinetic scales in the fast solar wind turbulence}


\author{Sonny Lion, Olga Alexandrova and Arnaud Zaslavsky\affil{LESIA, Observatoire de Paris, PSL Research University, CNRS, Sorbonne Universités, UPMC Univ. Paris 06, Univ. Paris Diderot, Sorbonne Paris Cité}}
\email{sonny.lion@obspm.fr}

\begin{abstract}
In this paper we investigate spectral and phase coherence properties of magnetic fluctuations in the vicinity of the spectral transition from large, magnetohydrodynamic (MHD) to sub-ion scales using in-situ measurements of the Wind spacecraft in a fast stream. For the time interval investigated by Leamon et al. (1998) the phase-coherence analysis shows  the presence of sporadic quasi-parallel Alfv\'en Ion Cyclotron (AIC) waves as well as coherent structures in the form of large-amplitude, quasi-perpendicular Alfv\'en vortex-like structures and current sheets. These waves and structures importantly contribute to the observed power spectrum of magnetic fluctuations around ion scales; AIC waves contribute to the spectrum in a narrow frequency range whereas the coherent structures contribute to the spectrum over a wide frequency band from the inertial range to the sub-ion frequency range.
We conclude that a particular combination of waves and coherent structures determines the spectral shape of the magnetic field spectrum around ion scales. This phenomenon provides a possible explanation for a high variability of the magnetic power spectra around ion scales observed in the solar wind.

\end{abstract}

\bibliographystyle{apj}


\section{Introduction}
In usual hydrodynamical (HD) flow, a turbulence cascade develops between energy injection and energy dissipation scales and its spectrum $\sim k^{-5/3}$ can be described by the Kolmogorov's phenomenology \citep{Kolmogorov1941}.  
The coherent structures, responsible for intermittency, are filaments of vorticity which are localized in space but  cover all scales, from the energy injection scale, up to the dissipation scale $\ell_d$, i.e. their cross section is of the order of $\ell_d$ \citep{Frisch1995}. 
The situation is mostly the same for MHD turbulence, where the absence of characteristic scales gives rise to a  well defined power law behaviour of the turbulent fluctuations spectrum. Dissipative processes are in this case related to the plasma resistivity, and coherent structures, responsible for intermittency, are usually current sheets with the thickness of the order of $\ell_d$.  

However, in the solar wind, the resistivity as well as the viscosity are extremely low, and turbulence can develop down to characteristic scales of the plasma before being dissipated. The first range of characteristic scales encountered are the ion kinetic scales, such as ion cyclotron frequency $f_{ci}=q_iB_0/2\pi m_i$ ($B_0$ being the mean magnetic field), the  ion Larmor radius  $\rho_i=v_{th \perp i}/2\pi f_{ci}$,  with ion thermal speed $v_{th \perp i} = \sqrt{2k_BT_{\perp i}/m_i}$ ($T_{\perp i}$ being the ion temperature perpendicular to ${\bf B_0}$), and  the ion inertial length $\lambda_i = c/\omega_{pi}$ (the ion plasma frequency is defined as $\omega_{pi}=q_i\sqrt{n_i/\epsilon_0m_i}$, $\epsilon_0$ being the permittivity of free space). Observations show that the Kolmogorov-like cascade ends at these scales,  and the spectrum is observed to be steeper at smaller scales, exhibiting another power-law behavior in the so-called kinetic range (e.g. \citet{Alexandrova2013}). In between these two power-law regimes, lies the so-called transition range, around the ion characteristic scales. The spectral shape of this transition region is quite variable. It can sometimes be adequately fitted by a power-law (as was done, e.g. in \citep{Smith2006} or in \citep{Sahraoui2010}), but sometimes exhibits a non-power law smooth transition behavior, like observed by \citet{Bruno2014a}, for example,  or even positive slopes in the presence of quasi-monochromatic waves \citep{Jian2014}.

Several authors studied this transition range by introducing a break frequency $f_b$, defined as the upper-boundary of the Kolmogorov cascade. This frequency was usually determined to be the intersection of slopes obtained by independent linear fits of the lower and upper frequency parts of the spectrum (e.g. \citet{Bourouaine2012}). Furthermore, numerous studies attempted to correlate the break frequency $f_b$ with ion plasma characteristic scales in order to determine which physical process controls the spectral steepening. This includes, for example, the studies of \cite{Leamon2000, Markovskii2008, Perri2010, Bourouaine2012, Bruno2014a}. 
But these authors do not agree on which scale, if any,  is best correlated with $f_b$, and therefore which process governs the physics of the spectral steepening.

Our purpose here is to study which physical processes are at work around ion scales, how these processes influence the spectral shape at these scales and therefore why the transition region does sometimes exhibit a clear spectral break and sometimes not.
To do this, we chose to re-analyse a fast solar wind interval used in Fig.~1 of \cite{Leamon1998} 
wherein a sharp and well defined break was observed. 
We use the Morlet wavelet transform, 
which gives us  the possibility to have information on the local phase of the signal. 
Using this tool, we show that the clear spectral break at the frequency $f_b \simeq 0.4$~Hz results from the superimposition of (i) non-coherent and not-polarized fluctuations, (ii)  emissions of parallel propagating Alfv\'{e}n Ion Cyclotron (AIC) waves in a narrow frequency range around $f_b$ and (iii)  large amplitude coherent structures in form of Alfv\'en vortex-like structures and current sheets. These structures cover a very large range of scales in the inertial range, their smallest scale, or their characteristic size appears to be a few $\lambda_i$ or $\rho_i$, that is close to $f_b$ by Doppler shift. 
 
Our work also sheds light on the nature of intermittency in the solar wind. Previous studies showed the role of planar structure like current sheets, rotational discontinuities and shocks in solar wind intermittency \citep{Greco2012, Perri2012, Salem2009, Veltri1999}. Here, we point out that besides current sheets and rotational discontinuities, we find also the signature of vortex-like structures at ion scales indicating that our vision of solar wind intermittency was too restricted to planar structures ignoring the filamentary structures like vortices.

Our results clearly show, as well, that the physics of solar wind turbulence around ion scales is not governed by a single physical process.
Because the proportion of structures, waves and non-coherent fluctuations is not always the same and depends on local (wave instabilities) and non-local (convection of structures) phenomena, the spectral shape may vary from time to time. These results may help to explain why the break is not a permanent feature in the solar wind and also why there is no single characteristic scale which controls the spectral steepening at ion scales (as it is usually defined).

The paper is organized as follows: Section \ref{sec:Interval_overview} describes the spacecraft data and summarizes the plasma parameters involved, as well as the spectral properties of the selected interval; Section \ref{sec:phase_analysis} includes the identification and characterization of the most energetic and polarized magnetic fluctuations of the time interval. In Section \ref{sec:phase_filtering}, we detail the spectral contribution of waves, coherent structures and the rest of the signal by using wavelet coherency technique. Finally, Section \ref{sec:conclusion} summarizes our findings and brings our conclusions.


\section{Interval overview}
\label{sec:Interval_overview}

\begin{figure}
\begin{center}
\includegraphics[width=\simplesize]{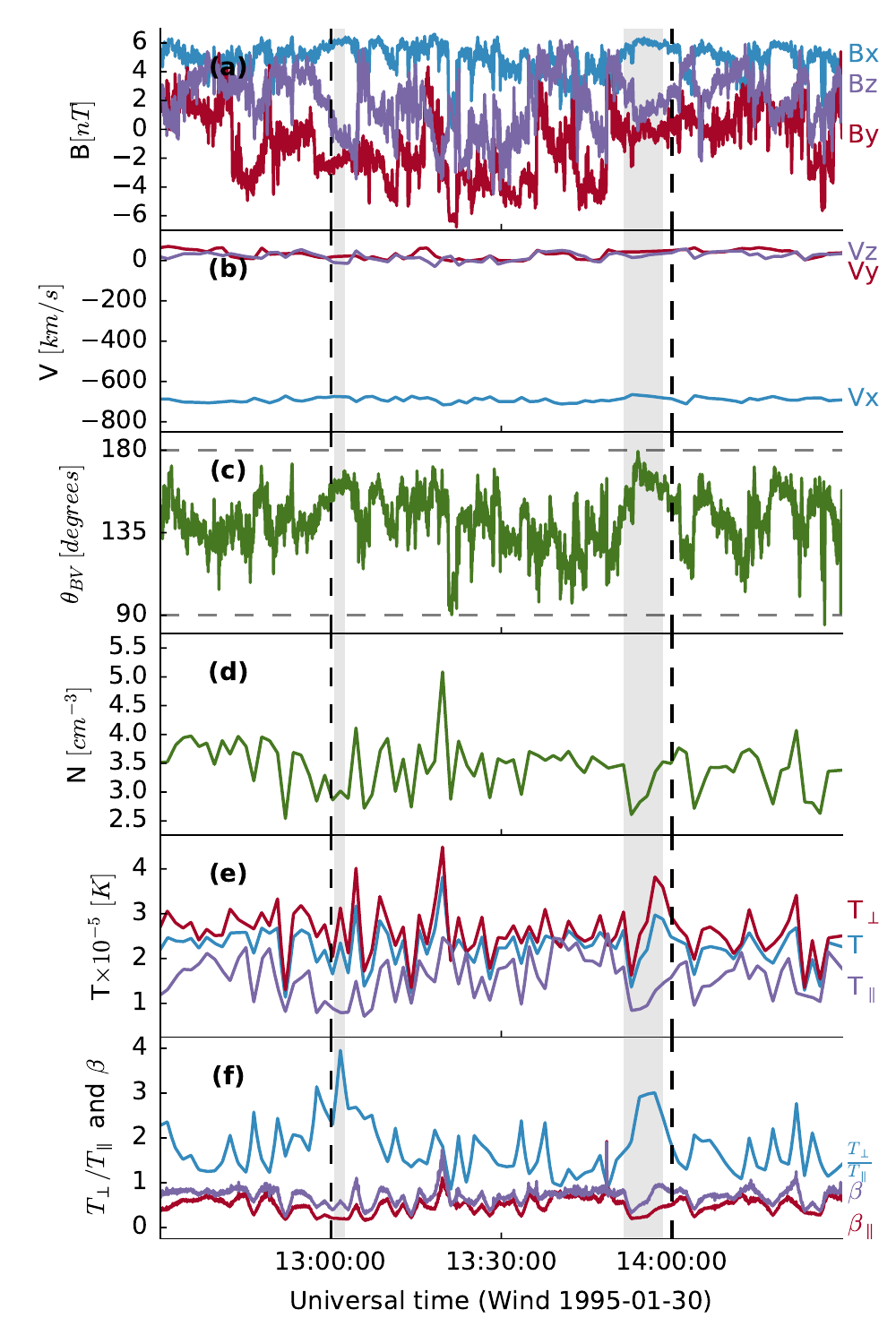}
\caption{Time series of solar wind data as measured by Wind spacecraft on January 30,1995. From the top down, the six panels show (a)~the vector magnetic field $\textbf{B}$, (b)~the vector proton velocity $\textbf{V}$, (c)~the angle between $\textbf{B}$ and $\textbf{V}$, (d)~proton density, (e)~proton total, parallel and perpendicular temperature, with  respect to the mean magnetic field (f)~plasma ion $ \beta $, $ \beta_\parallel $ and temperature anisotropy ($T_\perp/T_\parallel$). Grey filled bands represent two areas where AIC waves are present (as shown in section \ref{sec:phase_analysis}). The central interval between two dashed lines corresponds to Figure 1 of \citet{Leamon1998} and is used here for the spectral analysis, whereas the entire interval is used for the rest of the study.}
\label{fig1}
\end{center}
\end{figure}

We consider here a two hours interval [12:30:00-14:30:00]~UTC on the 1995/01/30 of measurements of the local magnetic field vector obtained by the MFI instrument \citep{Lepping1995} onboard the {\it Wind} spacecraft with a resolution of 184 ms; we use also  the 92 s resolution proton data from the SWE instrument \citep{Ogilvie1995}. Since these magnetic field and proton measurements are not evenly spaced in time and have time  gaps, we interpolated all the data to the resolution of 184~ms for MFI and 92~s for SWE. All vector data are given in the GSE reference frame.

Figure~\ref{fig1} shows the plasma parameters in the studied time interval, in the following order: the three components of the magnetic field $ \textbf{B} $, the three components of the velocity vector $ \textbf{V} $, the angle $ \theta_{BV} $ between $ \textbf{B} $ and $ \textbf{V} $, the proton density $ N $, the total proton temperature $ T $ (blue line), perpendicular $ T_\perp $ (red line) and parallel $ T_\parallel $  (purple line) proton temperatures ($\|$/$\perp$ with respect to the mean magnetic field $ \textbf{B}_0 $), the plasma parameters $ \beta =nkT/(B^2/2\mu_0)$ (purple line) and $ \beta_\parallel=nkT_{\|}/(B^2/2\mu_0) $ (red line), and the temperature anisotropy $ T_\perp / T_\parallel$ (blue line). The central interval between 13:00:00  and 14:00:00~UTC delimited by dashed lines is the interval used in Fig.1 of \citet{Leamon1998}. Two areas of interest discussed below are indicated by grey filled bands.

Table~\ref{tbl:wind_plasma_parameters} summarizes the parameters of the plasma in the central interval. In particular, it can be noted that the mean velocity corresponds well to the fast wind stream $ |\textbf{V}_0|  = (691 \pm 12)$~ km/s. Moreover the field-to-flow angle is $ \theta_{BV} = (143 \pm 17)^{\circ} $, which implies that ${\bf B}_0 $ nearly follows the Parker spiral \citep{Parker1958} and is directed sunward. The complete interval presents similar characteristics.

\begin{table}
\centering
\begin{tabular}{|c|c|}
 \hline
 $B$ (nT) & $      6.3 \pm      0.2$ $(0.8, -0.2, 0.3)$ \\
 $V$ (km/s) & $    691 \pm     12$  $(-0.998, 0.045, 0.032)$\\
 $\theta_{BV} (^{\circ})$ & $    143 \pm     17$ \\
 $N$(cm$^{-3}$) & $      3.4 \pm      0.4$ \\
 $T$ (K) & $ (22.8 \pm  3.9)\times 10^{4}  $ \\
 $T_\perp/T_\parallel$ & $      1.8 \pm      0.6$ \\
 $\beta \mbox{ ; } \beta_\parallel$ & $      0.7 \pm      0.2 \mbox{ ; } 0.5 \pm 0.2$ \\
 $\rho_i$ (km) & $108 \pm 10$ \\
 $\lambda_i$ (km) & $124 \pm 6$ \\
 $f_{ci}$ (Hz) & $0.096 \pm 0.003$ \\
 \hline
\end{tabular}
\caption{Average plasma parameters between 13:00:00 and 14:00:00~UTC, on 30 January 1995, as measured by Wind spacecraft. Unit vectors are given in brackets in GSE reference frame.}
\label{tbl:wind_plasma_parameters}
\end{table}


\begin{figure}
\begin{center}
\includegraphics[width=\simplesize]{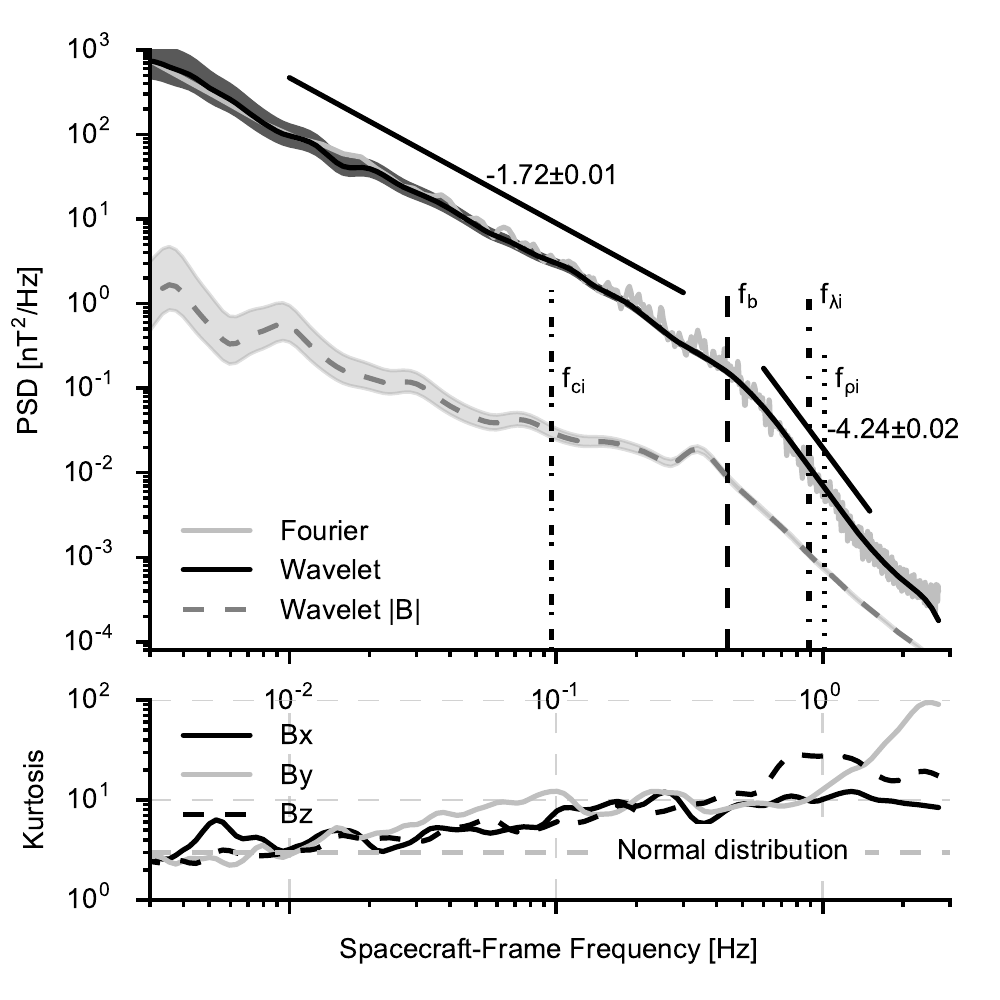}
\caption{Upper panel: Total magnetic field Power Spectral Density (solid lines) as a function of frequency in the spacecraft frame  as measured by Wind  on January 30,1995, from 13:00 to 14:00~UTC, computed with FFT (grey) and Morlet wavelet (black) algorithms. The spectrum of compressible magnetic fluctuations (grey dashed line), computed from the magnetic field modulus with Morlet wavelets, $PSD(|{\bf B}|)$. Straight lines show power-law fits. Vertical lines indicate the break frequency $f_b$ (dashed), the ion-cyclotron frequency $f_ {ci}$ (dashed-dotted, left) as well as the inertial length $f_{\lambda i}$ (dashed-dotted, right) and the Larmor radius $ f_{\rho i}$ (dotted) Doppler-shifted to spacecraft frequency using the Taylor hypothesis \citep{Taylor1938}.
Lower panel: kurtosis for each component of the magnetic field computed from wavelet coefficients. The horizontal dashed line indicates a value of $3$ expected for a normal distribution. The bump in $PSD(|{\bf B}|)$ at 0.33~Hz and the oscillations of the kurtosis of $B_x$ and $B_y$ are due to the spinning frequency of Wind spacecraft.
}
\label{fig2}
\end{center}
\end{figure}

The spectrum of the central interval (between 13:00:00 and 14:00:00~UTC) is presented in the upper panel of Figure~\ref{fig2} as a function of the frequency in the spacecraft frame.
The black curve shows the total PSD calculated with the continuous wavelet transform (see eq.~\eqref{total_psd} in Appendix~A), the filled area represents its $95\%$ confidence interval (eq.~\eqref{confidence_limits}), while the grey curve shows, for comparison, the total spectrum calculated with windowed Fourier transform by applying a pre-whitening and post-darkening as was done by e.g. \citet{Leamon1998, Bieber1993}.
The dashed line  shows the spectrum of the magnetic field modulus $ B = | \textbf{B} | $, which is used as a proxy for the compressible fluctuations. Again, the filled area represents the $95\%$ confidence interval.
One can notice that the confidence interval is negligible compared to the thickness of the curve for frequencies higher than $0.1$~Hz. For the other spectra of this paper, the confidence intervals are almost the same as in Figure~\ref{fig2}.

Spectra end at $ \sim 2.7$~Hz, the Nyquist frequency of the MFI data of this interval.  At frequencies $f > 1.6$~Hz, appears a flattening of the spectrum caused by the instrumental noise contribution.
The probe spin at $ 0.33$~Hz is not visible on the total spectrum, but appears in the modulus spectrum as an excess of energy around that frequency. The break frequency  
$f_b \simeq 0.44$~Hz, as well as characteristic plasma frequencies $f_ {ci}  \simeq 0.096$~Hz, $f_{\lambda i} = \frac{V}{2\pi \lambda_i} \simeq 0.89$~Hz and $ f_{\rho i} = \frac{V}{2\pi \rho_i} \simeq 1.0$~Hz are indicated by vertical lines. Note, that the ion characteristic scales cover, in this case, one decade in frequencies $\sim [0.1,1]$~Hz.

We also show power law fitting for frequency ranges $ [10^{- 2}, 0.3]$~Hz (inertial range) and $ [0.6,1.6]$~Hz (transition range), respectively $ f^{-1.72} $ and $ f^{-4.25} $. These results are similar to those obtained by \citet{Leamon1998}. We note that $f^{-4.25}$ is obtained for a very short frequency interval.
The break frequency $f_b$ is near the plasma characteristic frequencies but does not match any of them.

The lower panel of Figure~\ref{fig2} shows the kurtosis \citep{jones_scipy_2001, Zwillinger2000} for three components of the magnetic field in the GSE frame as a function of the frequency. 
Within the inertial range, the kurtosis of the three components increases with the frequency, as expected for an intermittent turbulent cascade. At ion scales, the kurtosis of the three components changes its behavior: it shows a plateau.  Such a plateau around ion scales have been already observed in the solar wind by \citet{Alexandrova2008a} and \citet{Wu2013}. Then at sub-ion scales, the kurtosis of the $B_y$ component  increases again, while for $B_x$ and $B_z$ the plateau continues. However, at these high frequencies the measurements are too close to the instrumental noise to give any firm conclusions. The ratio between the spectrum of compressible fluctuations and the total PSD, the so-called 
compressibility index (not shown here) increases with increasing frequency as one approaches the ionic scales as already observed by \citet{Alexandrova2008a, Salem2012, Kiyani2013}.


\section{Waves and structures identification}
\label{sec:phase_analysis}

\begin{figure*}
\begin{center}
\includegraphics[width=\doublesize]{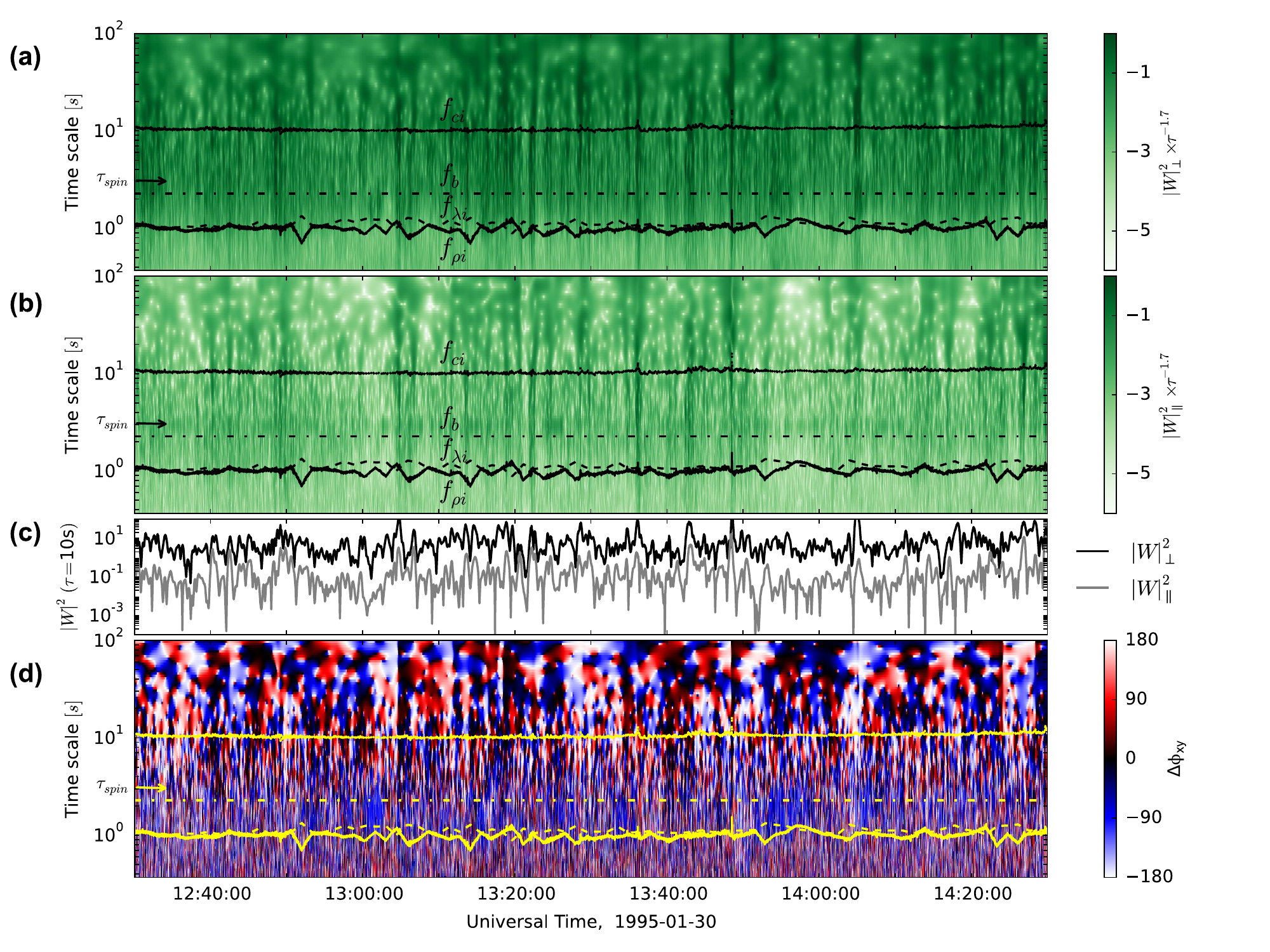}
\caption{Time evolution of magnetic field fluctuations: (a) compensated scalogram of the  energy of transverse fluctuations, $(|W_x(f,t)|^2+|W_y(f,t)|^2) \times f^{1.7}$ as a function of time and time-scales $\tau=1/f$, black solid and dash-dotted lines indicate $f_b$, $f_ {ci} $, $f_{\lambda i}$ and $ f_{\rho i}$, an arrow indicates the spin time-scale $\tau_{spin}=3$~s of the Wind spacecraft, (b) the same as (a) but for the parallel fluctuations $|W_z(f,t)|^2 \times f^{1.7}$ , (c) a cut of perpendicular (black) and parallel (gray) scalograms at $f= 0.1$~Hz ($\tau= 10$~s), (d) phase differences between $B_x$ and $B_y$ calculated using complex Morlet wavelets, $\Delta \Phi_{xy}(t,\tau)$,  which give a polarization map in the plane perpendicular to ${\bf B_0}$, as a function of time and time-scales.}
\label{fig3}
\end{center}
\end{figure*}

The spectra give a global vision of the properties of an interval but do not allow to distinguish between the different processes at work which can appear at different times. On the contrary, wavelet spectrograms (or scalograms) allow to follow the energy evolution of the magnetic field fluctuations in the time-frequency (or time-scale) space \citep{Farge1992, Farge2015}.


We use thereafter scalograms compensated by the inertial range spectrum observed here $\propto  f^{-1.7}$ (see Figure~\ref{fig2}). This compensation allows to see better variations of the magnetic energy relative to the background.

We use also the flow-field reference frame defined as follows:
\begin{equation}
\begin{array}{ll}
\textbf{e}_z&=\textbf{B}_0/|\textbf{B}_0| \\
\textbf{e}_x&=\textbf{e}_z \times \textbf{V}_0/|\textbf{V}_0| \\
\textbf{e}_y&=\textbf{e}_z \times \textbf{e}_x
\end{array}
\end{equation}

Compensated scalograms of perpendicular and parallel energy of magnetic fluctuations, $(|W_x(f,t)|^2+|W_y(f,t)|^2) \times f^{1.7}$ and $|W_z(f,t)|^2 \times f^{1.7}$, are respectively presented in Figure~\ref{fig3}(a) and \ref{fig3}(b) as functions of time and time-scale $ \tau = 1 / f $. Figure~\ref{fig3}(c) shows a cut of the parallel and perpendicular scalogram at $ 0.1$~Hz ($ 10$~s). Finally Figure~\ref{fig3}(d) shows the polarization map in the plane perpendicular to $ \textbf{B}_0 $ built with the phase differences $ \Delta \Phi_{xy} (f, t) $ between $B_x$ and $B_y$ (see  eq.~\ref{eq:phase_diff}, Appendix A), in the way of a scalogram where blue and red areas represent the left-handed and right-handed polarized fluctuations, respectively,  and black/white is used for the linear polarization.

The scalograms show clearly high-energy events that span almost all frequencies, for example around 13:37:00~UTC. This coupling over many scales is an intrinsic property of coherent structures \citep{Frisch1995, Alexandrova2013}. The scalogram of the parallel component also reveals an excess of energy around $ 0.33$~Hz ($ 3$~s), signature of the probe spin. This excess is only visible in the parallel fluctuations because, as shown in the scalograms, the perpendicular components are more energetic than the parallel one. The magnetic fluctuations energy is lower at the edges of the central interval  (around 13:00 and 14:00~UTC) mainly for the parallel component. These properties are especially visible in the Figure~\ref{fig3}c and we see that $10 \leq \frac{|W_\perp|^2}{|W_\parallel|^2} \leq 100$ during the whole interval.
 
Finally the polarization map (Figure~\ref{fig3}, bottom panel) shows that the phase differences highly fluctuate and seem, at first sight, distributed almost randomly. However we observe at the beginning (13:00:30 to 13:02:30~UTC) and at the end (13:51:30 to 13:58:30~UTC)  of the central interval two events indicated by gray bands on Figure~\ref{fig1}, during which the polarization remains around $ -90^{\circ} $ at frequencies around $ f_b $ (horizontal yellow dash-dotted line). These events correspond to left-handed polarized waves in the magnetic field reference frame and appear when the turbulence background level and especially $W_\parallel^2$ are the lowest (see Figure 3c). The first wave lasts 2 minutes (13:00:30 to 13:02:30~UTC) with a frequency between $ 0.4 $ and $ 1$~Hz (or time scales $\tau\in [1,2.5]$~s). The local spectrum, calculated over these 2 minutes, has a bump in this frequency range (not shown). Here, $\textbf{B}$ and $\textbf{V}$ are almost parallel ($\theta_{BV} \simeq 160^{\circ} $).
Using the minimum variance analysis \citep{Sonnerup1998} and assuming that the wave vector $ \textbf{k} $ is in the minimum variance direction $ \textbf{e}_{min} $, we obtain that the angle $ \theta_{kB} \simeq 14^{\circ} $ i.e. that $ \textbf{k} $ and $ \textbf{B}_0 $ are almost parallel. The second wave which lasts 7 minutes (13:51:30 to 13:58:30~UTC) has similar properties.

\begin{figure*}
\begin{center}
\includegraphics[width=\doublesize]{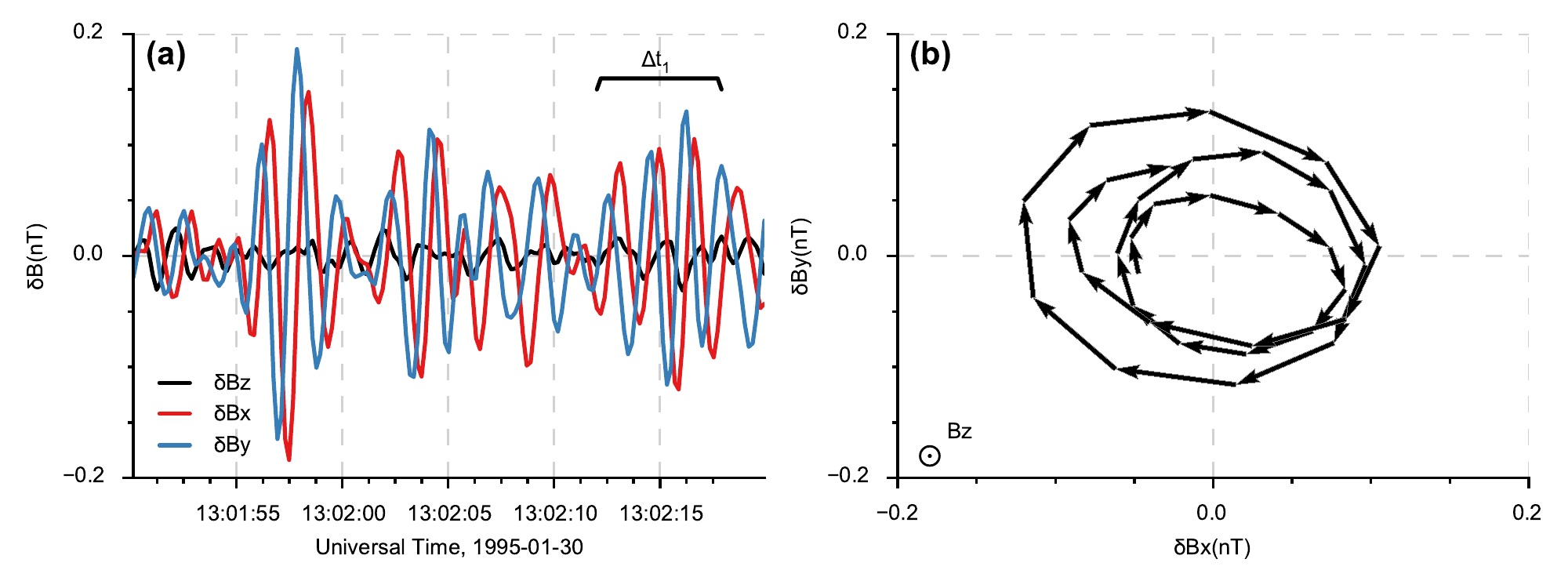}
\caption{An example (30 seconds zoom) of coherent AIC wave observed by Wind on 1995 January 30 during the analyzed time interval. Left panel: magnetic field fluctuations within $[0.4,1]$~Hz frequency range in the local field aligned frame, with  $\textbf{B}_0$ along ${\bf e}_z$. Right panel: left-handed polarization in the plane perpendicular to $\textbf{B}_0$ represented as a hodogram.}
\label{fig4}
\end{center}
\end{figure*}

Figure~\ref{fig4} provides a zoom on magnetic field fluctuations of the 3 components filtered between $f_{min} = 0.4 $ and $f_{max} = 1$~Hz for 30~seconds time interval within the first left-handed Alfv\'{e}n wave emission. 
Fluctuations are defined as:
\begin{equation}
\delta B_x = L_{f_{max}}(B_x) - L_{f_{min}}(B_x)
\label{smooth}
\end{equation}
with $L_{fs}$ the moving average at frequency $ f_s $ (window of size $1/f_s$). The left panel shows the time evolution of the fluctuations,  while the right panel shows the hodogram of fluctuations in the plane perpendicular to $ \textbf{B}_0 $ in the time interval $ \Delta t_1 $ (indicated in the left panel). Fluctuations of the perpendicular components are of the same intensity and with a phase shift of $ - \pi / 2 $ which results in a left handed (with respect to $\textbf{B}_0$) almost circular rotation in the hodogram. Maximum and medium eigenvalues of the minimum variance matrix ​​are almost equal ($ \lambda_{med} / \lambda_{max} \sim 0.8$), while the minimum eigenvalue is very small compared to the maximum one ($ \lambda_{min} / \lambda_{max} \sim 10^{- 3} $), i.e. $\textbf{e}_{min}$ is well defined.

\begin{figure*}
\begin{center}
\includegraphics[width=\doublesize]{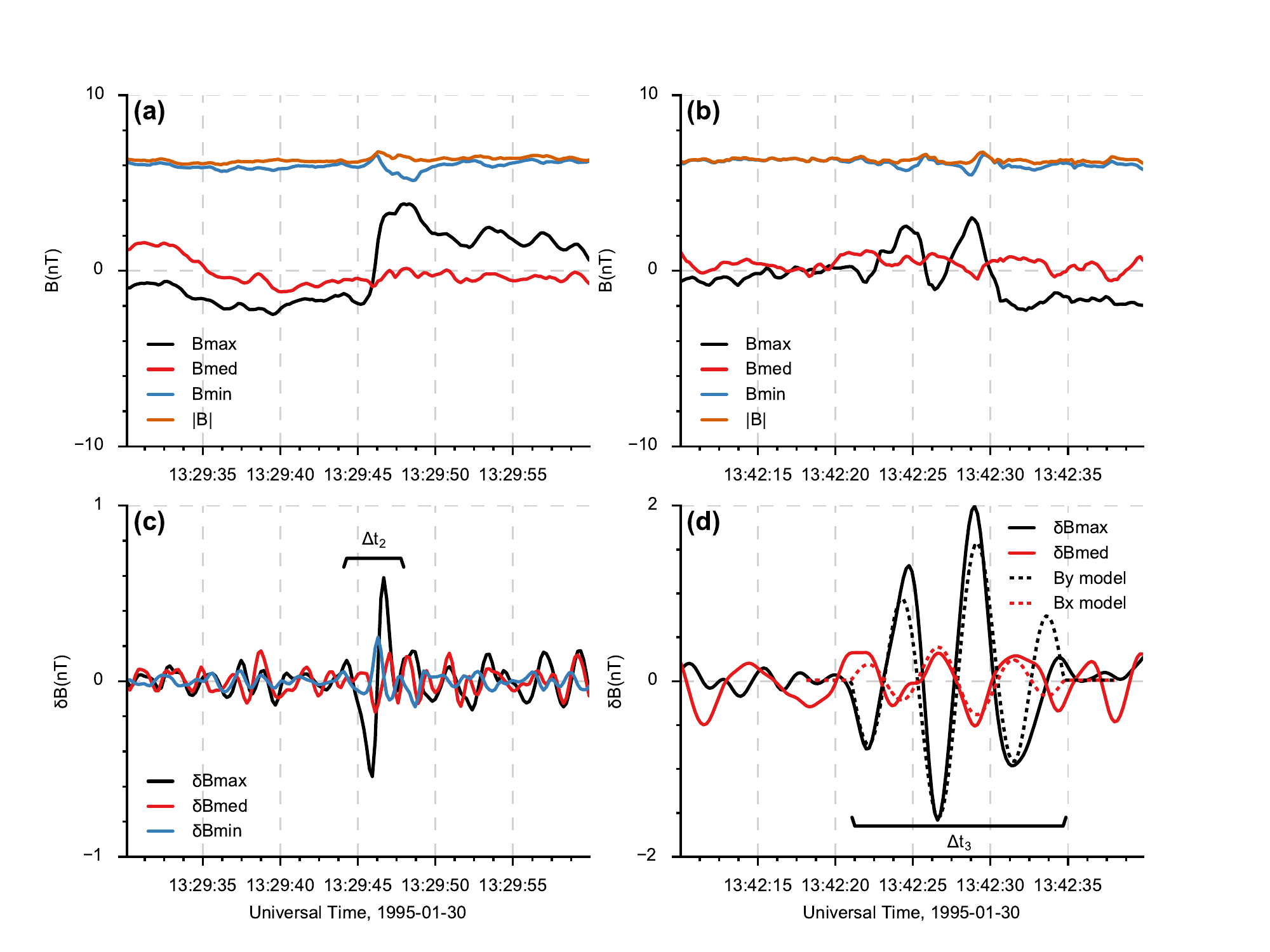}
\caption{Two examples of coherent energetic events presented in the minimum variance frame calculated over 30 seconds time intervals shown here. Left panels: current sheet detected at 13:29:46 (a) raw data, (c) fluctuations defined by eq.~\eqref{smooth} in the $[0.4,1.0]$~Hz frequency range. Duration $\Delta t_2 \simeq 4$~s. Right panels: signatures of an Alfv\'{e}n vortex-like structure at 13:42:27 (b) raw data, (d) fluctuations defined in the $[0.1,0.4]$~Hz frequency range superposed to the monopole Alfv\'{e}n vortex model from \citet{Petviashvili1992}. Duration $\Delta t_3 \simeq 14$~s.} 
\label{fig5}
\end{center}
\end{figure*}

In addition, as shown in Figure~\ref{fig1}, during the two waves events (grey filled bands), the temperature anisotropy increases while $ \beta_\parallel $ decreases. Here $ T_\perp/T_\parallel \simeq 3.5 $ and $ \beta_\parallel \simeq 0.2 $ which is compatible with Alfv\'{e}n ion cyclotron (AIC) instability \citep{Gary1994} and in accordance with the growth rates between $ \gamma_{max} = 10^{- 3} \omega_{ci} $ and $ \gamma_{max} = 10^{- 1} \omega_{ci} $ \citep{Hellinger2006}. The generation of these low frequency waves by AIC instability due to a temperature anisotropy seems a plausible scenario. Our analysis thus suggests that these waves are parallel ion cyclotron Alfv\'{e}n waves.

Figure~\ref{fig5} shows two examples of magnetic fluctuations around the high-energy events that span a large range of scales in the scalograms of Figure~\ref{fig3}(a) and Figure~\ref{fig3}(b) (at 13:29:45 and 13:42:25~UTC). Note that these are only two examples among others with similar characteristics and they were chosen primarily because of their well-defined shapes. The upper panels show magnetic raw data projected to the minimum variance frame whereas the lower panels show the magnetic fluctuations ${\bf \delta B}$ (in the same reference frame). For panel (c), the fluctuations are defined (eq.~\eqref{smooth}) between $f_{min} = 0.4 $ and $f_{max} = 1.0 $~Hz;  and for panel (d), between $f_{min} = 0.1 $ and $f_{max} = 0.4$~Hz, which represents respectively the frequencies (or scales) covered by the events. 

One can see that the magnetic data in left panels correspond to a current sheet 
with an amplitude of $\delta B/B_0\sim0.75 $. The angle $\theta_{BV}=140^{\circ}$ is oblique,  ${\bf B_0}$ and 
${\bf e}_{min}$ are almost parallel ($\theta_{emin,B}=4^{\circ} $) whereas ${\bf B_0}$ and $\textbf{e}_{max}$ are perpendicular ($\theta_{emax,B}=90^{\circ}$). It seems that the current sheet is aligned with the mean magnetic field and the largest gradient is along the perpendicular direction. With the fluctuations plot we estimate an upper-bound for the temporal thickness of the current sheet $\Delta t_2 \sim 4$~s (or $1$~s peak to peak) considered here as a characteristic scale of the structure, which is the same order of magnitude as $f_b^{-1}\simeq 2.3$~s.
Assuming that this structure is convected by the solar wind, one can estimate its size using the projection of the velocity on the maximum variance axis $ V_{emax}=|\textbf{V}_0 \cdot \textbf{e}_{max}|\sim 300$~km/s. Therefore, the scale of the principal gradient (from peak to peak) is 300~km, or $2.4\lambda_i$ ($2.7\rho_i$); the scale corresponding to $\Delta t_2$ is  $1200$~km, or $10\lambda_i$ ($11\rho_i$), that can be considered as a scale where the current sheet affects  the surrounding plasma.

Figure~\ref{fig5}, right panels, represents magnetic fluctuations,  which look like a wave packet of  high amplitude ($\delta B/B_0\sim0.60 $). The principal fluctuations are in the plane perpendicular to ${\bf B_0}$ ($\theta_{emax,B}=93^{\circ}$, $\theta_{emed,B}=87^{\circ}$, $\theta_{emin,B}=5^{\circ}$) and the field-to-flow angle is oblique $\theta_{BV}=140^{\circ}$, as in the case of the current sheet example. 
Different models can explain this kind of fluctuations such as envelope soliton models \citep{Buti2000, Ovenden1983} or the Alfv\'{e}n vortex model \citep{Petviashvili1992}. 

Envelope soliton models describe fluctuations with $k_{\|} \gg k_{\perp}$. These wavevectors can be observed when  ${\bf B}$ and ${\bf V}$ are aligned ($\theta_{BV} \sim 0$), as in the case of the AIC waves described above. 
Here, we are in an oblique configuration, $\theta_{BV} = 140^{\circ}$ (or $40^{\circ}$), and, 
at the same time, the amplitudes of fluctuations are much higher than in the aligned case ($\theta_{BV} \sim 0$). 
If the solar wind turbulence is composed of a 2D component ($k_\perp \gg k_{\|}$) and a slab component ($k_{\|} \gg k_{\perp}$), as was suggested by \citet{Matthaeus1990}, in oblique $\theta_{BV}$ configuration, we will observe projections of these two components on the solar wind flow direction ${\bf V}$. However, as \citet{Horbury2008} and \citet{Wang2016} show, the 2D component has high amplitudes of fluctuations, and the slab component has much lower amplitudes (at $f<1$~Hz). Therefore, the projection of the 2D component will dominate the projection of the slab for oblique $\theta_{BV}$ angles. 
 So, it seems that the high-amplitude fluctuations we observe here under oblique $\theta_{BV}$ configuration have $k_\perp \gg k_{\|}$ and it is reasonable to consider the Alfv\'{e}n vortex model. 
  Moreover, similar fluctuations at almost the same scales have already been observed by four Cluster satellites in the Earth magnetosheath and have been interpreted as Alfv\'{e}n vortices \citep{Alexandrova2006}.
 
 Let us now verify whether the Alfv\'{e}n vortex model \citep{Petviashvili1992} can explain our observation. The magnetic field components of the Alfv\'{e}n vortex can be derived from the vector potential $\textbf{A}$ given by \citet{Alexandrova2008b}:
\begin{equation}
\begin{array}{ll}
\left\{
\begin{array}{ll}
B_x(x, y)&=y \left(\frac{2 \alpha x J_2(k r)}{r^2 J_0(a k)}-\frac{A_0 k J_1(k r)}{r}\right) \mbox{, } r < a \\
B_x(x, y)&=-\frac{2 a^2 \alpha x y}{r^4} \mbox{, } r \ge a
\end{array}
\right.\\
\\
\left\{
\begin{array}{ll}
B_y(x, y)&=\frac{2 \alpha x^2 J_0(k r)}{r^2 J_0(a k)} - \frac{2 \alpha (x^2-y^2) J_1(k r)}{k r^3 J_0(a k)} \\
&-\alpha + \frac{A_0 k x J_1(k r)}{r} \mbox{, } r < a \\
B_y(x, y)&= \frac{a^2 \alpha (x^2-y^2)}{r^4} \mbox{, } r \ge a
\end{array}
\right.
\end{array}
\end{equation}
Here $A_0$ is a constant amplitude, $J_i$ is the Bessel function of ith order, $r=\sqrt{x^2+y^2}$ is the radial variable in the plane of the vortex, $\alpha=\tan \gamma$ with $\gamma$ the angle between the normal to the plane $(x,y)$ and $\textbf{B}_0$. The vortex radius $a$, represents the radius of the circle where the fluctuations are concentrated. To ensure the continuity of the magnetic field components at $r=a$, $k$ is chosen to be one of the root $ j_{1,l}$ of $J_1$.  The comparison is done in Figure~\ref{fig5}(d), for $k=j_{1,3} \simeq 10.17$, $A_0=0.27$ in normalized units and $\alpha=0$, i.e. the Alfv\'en vortex has a monopole topology and it is static in plasma frame. This fitting corresponds to the trajectory of the satellite  across the center of the vortex, with a small angle of $13^{\circ}$ to the direction of the intermediate variance ${\bf e}_{med}$ (or $x$-axis of the vortex model). 
 
 One can see  that  the monopole Alfv\'en vortex model (dashed lines in Figure~\ref{fig5}(d)) fits well the observations (solid lines).  The small deviations can come from (i) the facts that the frequency band chosen to define the observed fluctuations (by eq.~(2)) is slightly larger  than the scales covered by the structure in the scalograms of Figure 3; (ii) a superposition of the neighboring events on the studied vortex-like structure. 

We estimate the radius of the vortex $a$ (that is  a half of the extension of magnetic fluctuations which are fitted to the vortex model)  to be $\Delta t_3/2 \sim 7$~s. The scale of the strongest gradient within the vortex (scale of the central field-aligned current filament) is of the order of $\tau =2$~s. Thus, the temporal scales of the vortex are around the break scale $f_b^{-1}\simeq 2.3$~s.  
Using the projection of the solar wind velocity to the vortex plane $(x,y)=({\bf e}_{max},{\bf e}_{med})$, $450$~km/s, we obtain the vortex radius  $a\sim 3\times 10^{3}$~km or $24\lambda_i$ and $27\rho_i$;  and the scale of the strongest gradient  $\ell\sim 900$~km or $7\lambda_i$ and $8\rho_i$. 



However, in order to have confidence in the interpretation in terms of the Alfv\'en vortex, a multi-satellite analysis should be performed. This will be a subject of our future work. Here, we focus on the question how the observed events as AIC waves and coherent structures close to $f_b$ influence the spectrum at ion scales.


\section{Influence of coherent events on the turbulent spectrum}
\label{sec:phase_filtering}

To determine how coherent processes as quasi-monochromatic waves and coherent structures affect the turbulent spectrum around ion scales, we propose to separate coherent events from the rest of the signal in order to estimate their respective contribution to the total spectrum.  We will separate (or filter) our data on the basis of the level of coherency, i.e. phase coupling,  between two magnetic field components. For this, we use
the wavelet coherence technique \citep{Grinsted2004}, applied previously in neuroscience and for geophysical time series to examine relationships in time frequency space between two time series. The details are given in Appendix~\ref{sec:phase_filtering_appendix}.

As we have just seen, ion cyclotron waves are plane waves propagating parallel to the mean magnetic field with a phase coupling between the two perpendicular components $B_x$ and $B_y$ as shown in Figure~\ref{fig3}(d) and Figure~\ref{fig4}. Coherent structures appear in perpendicular, Figure~\ref{fig3}(a), and parallel scalograms, Figure~\ref{fig3}(b). As a consequence, the study of the phase coherence in the plane perpendicular to ${\bf B_0}$ can reveal the presence of waves, while the same technique applied to a parallel and a perpendicular components should highlight coherent structures.

\begin{figure*}
\begin{center}
\includegraphics[width=\doublesize]{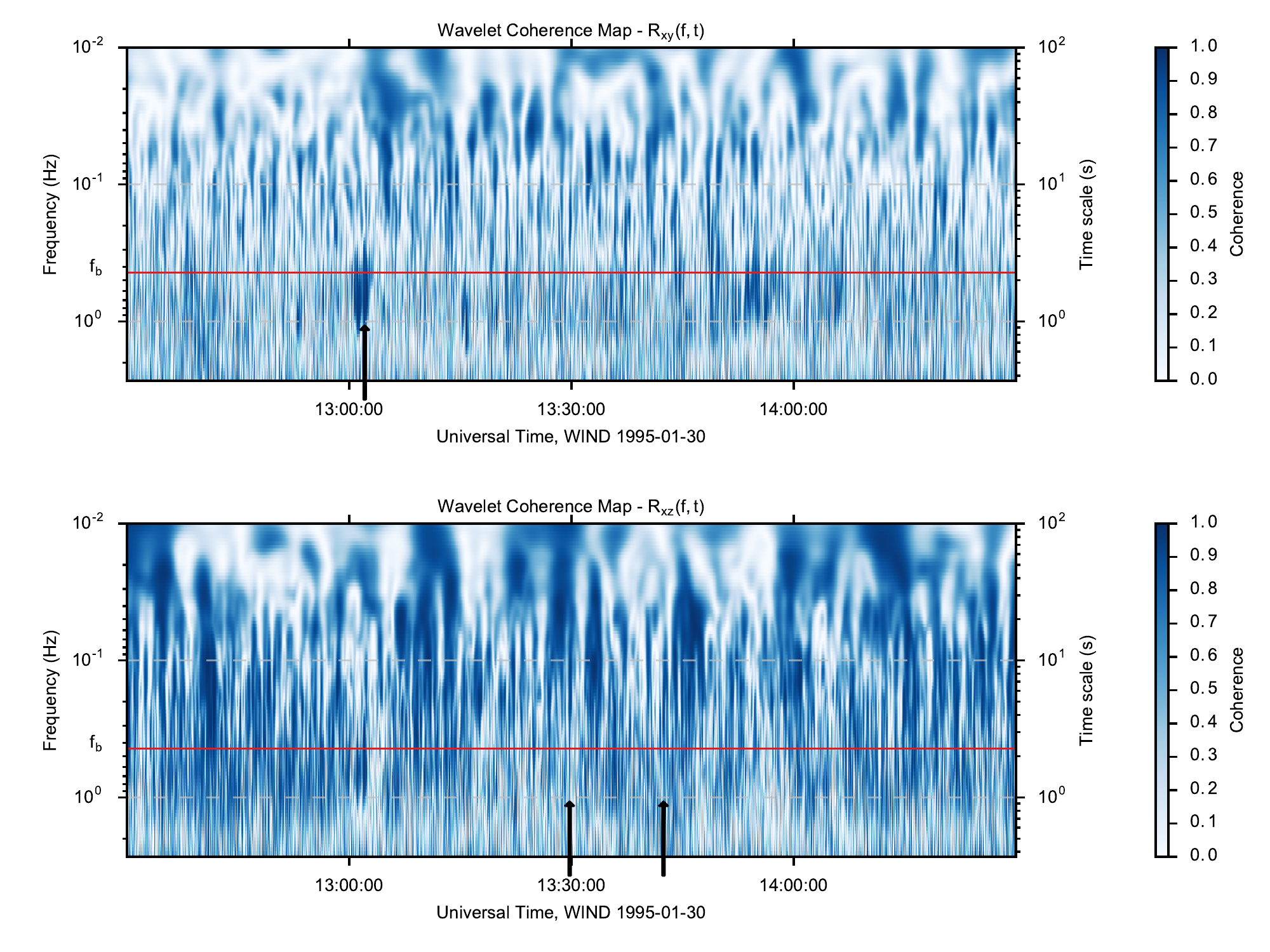}
\caption{Maps of phase coherency $R_{ij}(t,f)$ between two magnetic field components $B_i$ and $B_j$  as function of time and frequency. The red lines show the break frequency $f_b$. Upper panel: coherence between $B_x$ and $B_y$ components. Lower panel: coherence between $B_x$ and $B_z$ components.}
\label{fig6}
\end{center}
\end{figure*}

To keep an equivalent number of events after filtering and obtain a spectrum with statistical properties close to the ones of Figure~\ref{fig2}, we consider for the rest of the study the total interval of Figure~\ref{fig1}, i.e. the central interval ($1$~h) plus $ 30$~min on each side.

Figure~\ref{fig6} represents the phase coherency $R_{ij}(f,t)$ (see eq.~\ref{eq:coherence_map}, Appendix B) between $B_i$ and $B_j$ as a function of time and frequency. The top panel of Figure~\ref{fig6} corresponds to $R_{xy}(f,t)$, while the bottom panel corresponds to $R_{xz}(f,t)$.  Low levels of coherency ($R_{ij}(f,t)$ close to zero) correspond to light areas at these maps and high levels  of coherency ($R_{ij}(f,t)$ close to one) are dark areas: we will call them coherent areas.

For $R_{xy}(f,t)$, coherent areas are essentially in the frequency range close to or higher than $f_b$, whereas for $R_{xz}(f,t)$ coherent areas extend over almost all the frequencies and end on frequencies close to or above $f_b$.

\begin{figure}
\begin{center}
\includegraphics[width=\simplesize]{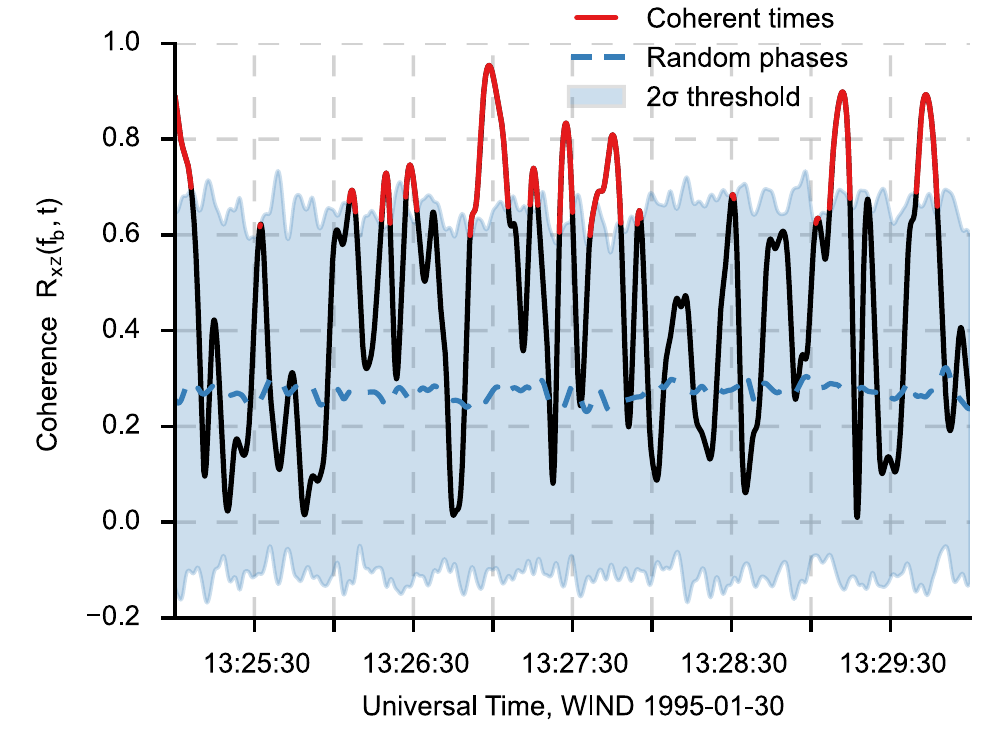}
\caption{Cut of $R_{xz}(f,t)$, shown in the lower panel of  Figure~\ref{fig6} at $f_b$. Here, red color corresponds to $R_{xz}(f_b,t)$ over the threshold (we refer to the corresponding times are coherent times); black corresponds to non-coherent times; dashed blue line shows the mean value (over 100 realizations) of $R_{xz}(f_b,t)$, but for the random phase surrogate signals; the filled ares shows the two times standard deviation of the surrogate data.}
\label{fig7}
\end{center}
\end{figure}

The filtering is done, using the coherence maps of Figure~\ref{fig6}, by selecting coherent areas above a threshold $R_{ij}^{threshold}$ (see eq.~\ref{eq:coherence_threshold} in Appendix B) at the break frequency $f_b$. 


To help the reader to better visualize how the selection is made, we show in Figure~\ref{fig7} a cut of the coherency map $R_{xz}(f, t)$ between 13:25:00 and 13:30:00~UTC at $f=f_b$. The $R_{xz}^{threshold}$ is given by the blue filled area. The sets of coherent and non-coherent times  correspond to $R_{xz}(f_b, t)$  above the threshold (red  line) and  below the threshold  (black  line) respectively. For comparison, the dashed line shows the average coherency over the different random signal realizations $\bar{R}^{random}_{xz}(f_b, t)$. Then coherent times are used to calculate individual spectra of coherent events, and non-coherent times, -- for the spectra of non-coherent fluctuations. For more information see Appendix B.

\begin{figure*}
\begin{center}
\includegraphics[width=\doublesize]{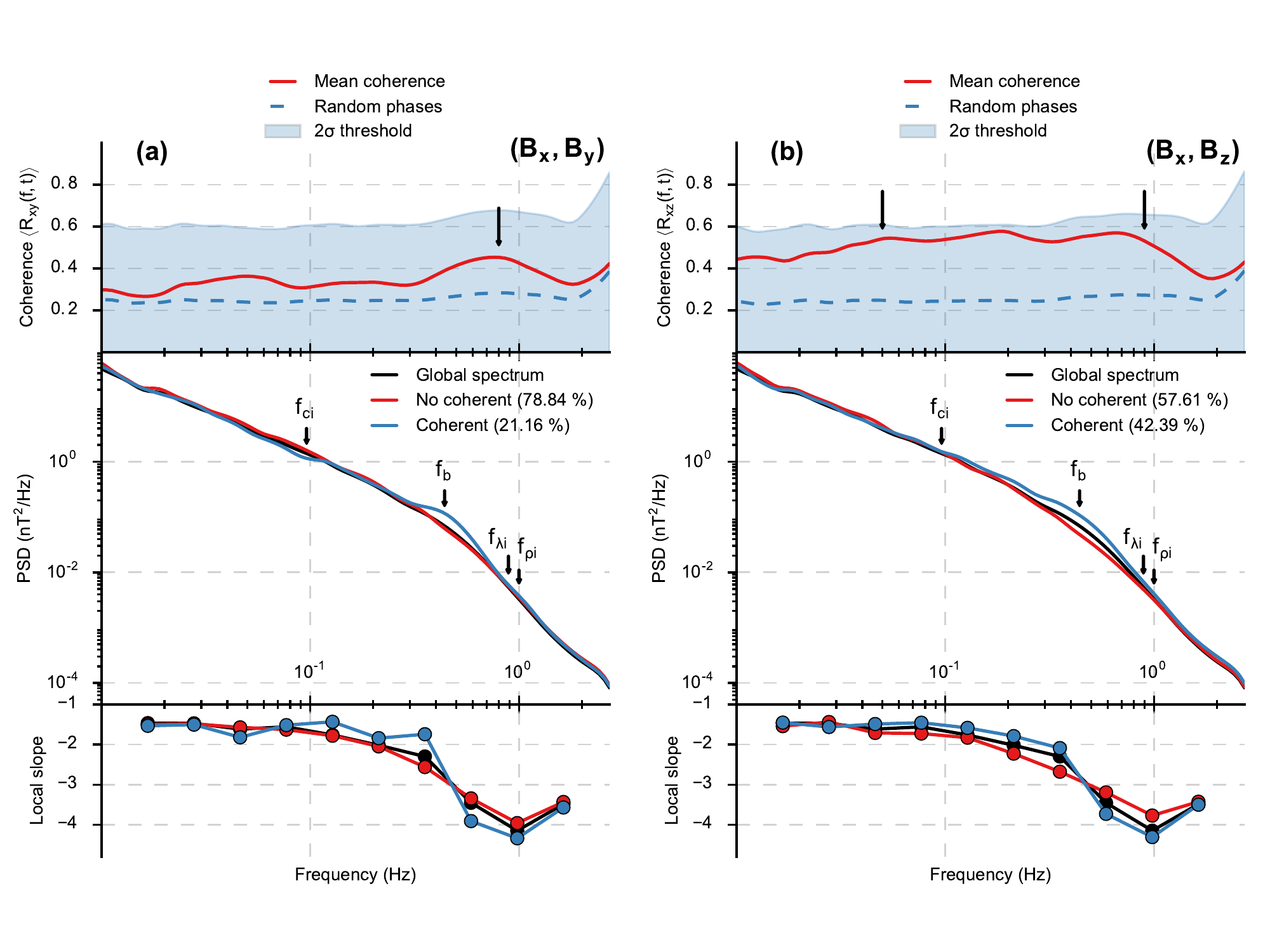}
\caption{Coherence and spectra as function of frequency for $(B_x,B_y)$ couple (left panels) and $(B_x,B_z)$ couple (right panels). Upper panels: averaged coherence over time (solid red), mean and two times standard deviation of surrogate data (blue dashed line and filled area). Center panels: spectrum over the whole interval (black), filtered spectra for the coherent (blue) and non-coherent (red) parts of the signal. Lower panels: local slopes corresponding to center panels spectra (same color code).}
\label{fig8}
\end{center}
\end{figure*}

The resulting individual spectra are shown in Figure~\ref{fig8} (middle panels) together with an average coherence  as a function of the frequency $\langle{\mathstrut R}_{ij}(f,t)\rangle_{t}$ (top panels) and the local slopes of the corresponding spectra\footnote{The local slope is computed as follow: we divide the logarithmic frequency axis into 10 intervals of equal length; the local slope is then obtained using a linear fit considering only the points within each interval.} (bottom panels). 

The average coherence gives information on the frequency localization of coherent events. Figure~\ref{fig8}(top, left) gives  this information for $B_x-B_y$ components  coupling:  $\langle{\mathstrut R}_{xy}(f,t)\rangle_{t}$ is plotted by solid red line and it is compared to the $\langle{\mathstrut R}_{xy}^{random}(f,t)\rangle_{t}$ (blue dashed lines), $\langle R_{xy}^{threshold}(f,t) \rangle_{t}$ is indicated by the blue filled area. One observes an increase of $\langle{\mathstrut R}_{xy}(f,t)\rangle_{t}$ just below $f_b\simeq 0.4$~Hz  and a maximum of $\langle{\mathstrut R}_{xy}(f,t)\rangle_{t}=0.46$ around $f=0.8$~Hz. This frequency range corresponds to the AIC waves. 

The average coherence  between $B_x$ and $B_z$, Figure~\ref{fig8}(top,right),
exhibits a plateau $\langle{\mathstrut R}_{xz}(f,t)\rangle_{t}\sim 0.5$ between $f=5 \times 10^{-2}$ and $f=0.9$~Hz (see two vertical black arrows). This increase of coherency over a large frequency range corresponds to the presence of coherent structures, mainly current sheets and vortex-like structures, over all these scales, as discussed in section 3. Note that, in this representation, it becomes clear that the smallest scale of the coherent structures is the local maximum at 0.7~Hz close to the right end of the plateau of the average coherency $\langle{\mathstrut R}_{xz}(f,t)\rangle_t$, i.e. in the vicinity of $f_b$.
Here, results close to $1.6$~Hz and beyond are to be interpreted with caution since we do not know the exact contribution of the noise. Finally, effects of instrumental filters on the signal phase and coherence for frequencies near the Nyquist frequency remain to be determined.

\begin{figure}
\begin{center}
\includegraphics[width=\simplesize]{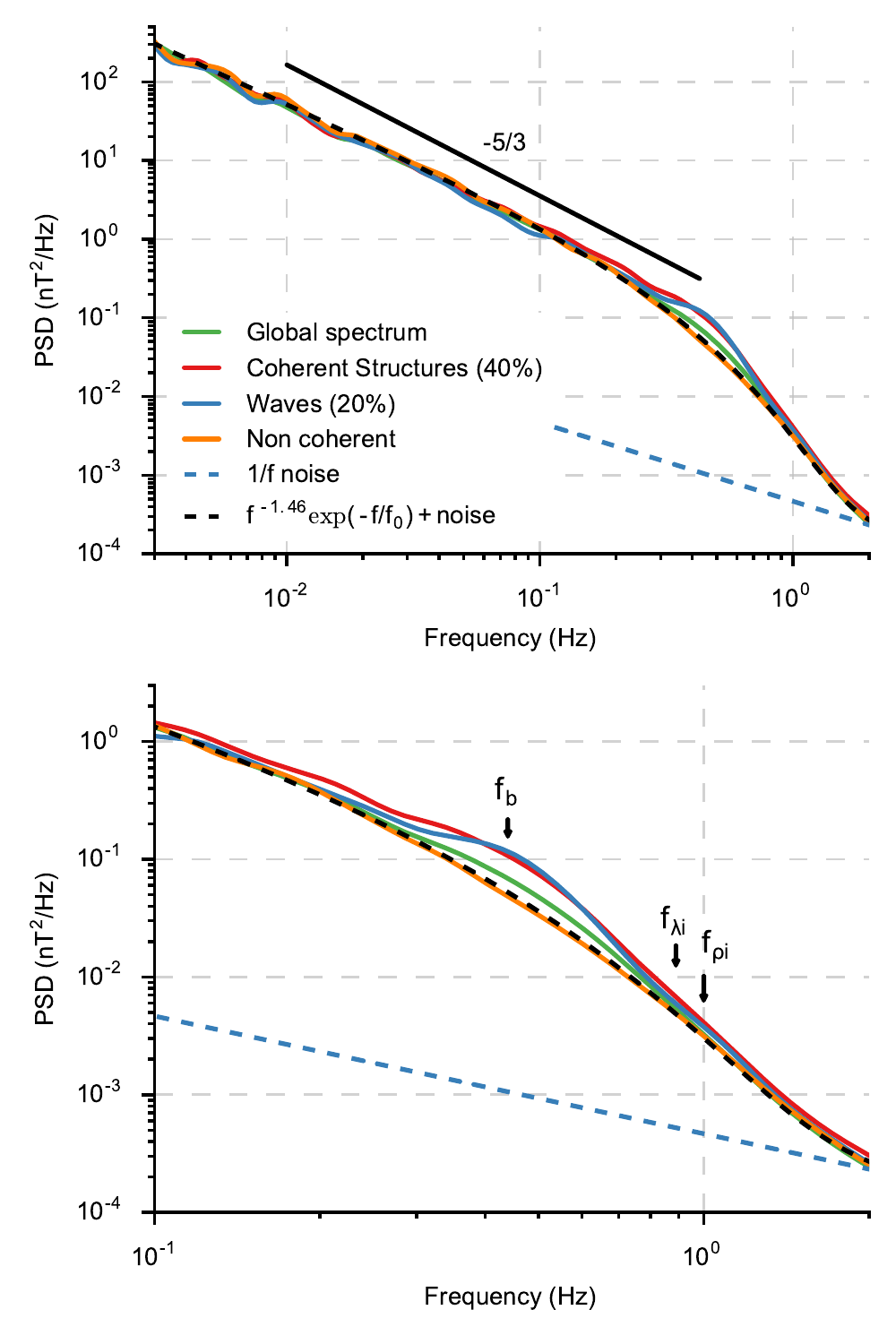}
\caption{Ionic transition spectrum decomposition. Upper panel: global spectrum (solid green); filtered spectra for wave signatures (solid blue), coherent structures (solid red) and non-coherent parts (solid orange); $1/f$ noise (dashed blue); exponential fit for the non coherent part (dashed black). Lower panel: zoom around the transition ($[0.1, 2]$~Hz, same color code as the upper panel).}
\label{fig9}
\end{center}
\end{figure}


Let us now consider  the individual spectra and compare them to the original spectrum for the whole time interval (or global spectrum), see central panels of Figure~\ref{fig8}. Here, the spectrum of coherent fluctuations $E^c$ is shown by a  blue solid line, the spectrum of the non-coherent part  $E^{nc}$  is in red and the global spectrum is in black.
Note that the global spectrum is the weighted average of coherent and non-coherent spectra\footnote{The spectrum over the whole time interval is $$E^{global}(f)=E^{c}(f)\frac{N_c}{N_{tot}} + E^{nc}(f)\frac{N_{nc}}{N_{tot}},$$ where $N_{c/nc/tot}$ are the number of points used to calculate coherent, non coherent and global spectra, respectively. }. 

Here, one observes that the spectra of non-coherent fluctuations do not exhibit any break. 
Also, we see  that the individual spectra of coherent events are higher than global and non-coherent spectra within the frequency ranges where the coherency increases (see top panels), i.e. around $f_b$, for $(Bx, By)$ and between $f_{ci}$ and $f_{\rho i}$ for $(Bx, Bz)$.
This result is consistent with Figure~\ref{fig3} scalograms, which show that coherent structures are among the most energetic events of the interval.

Left panels of Figure~\ref{fig8} show that waves affect the local spectrum creating a surplus of energy around their frequency, which results in a small bump (or a knee) around $f_b$ in the coherent spectrum (in blue). 

Right panels of Figure~\ref{fig8} show that the spectrum of coherent structures $E^c$ (in blue) has a  more pronounced break at $f_b$ than the global spectrum (in black). $E^c$  starts to deviates from the global spectrum  at $f\geq f_{ci}$ (see bottom panel).
In the inertial range ($f<f_{ci}$), the coherent spectrum is flatter than the non-coherent one with a local slope of 
$\alpha_c(0.077 \; Hz) \simeq -1.5$ and $\alpha_{nc}(0.077 \; Hz) \simeq -1.7$, 
respectively; whereas  around ion scales ($f\geq f_{ci}$), the local slope for the coherent  spectrum, $\alpha_c(0.6 \; Hz) \simeq -3.8$ is steeper than the non coherent one $\alpha_{nc}(0.6 \; Hz) \simeq -3.2$. The local slope of the global spectrum follows the slope of the coherent spectrum. Therefore we can deduce that the global scaling is imposed by the coherent part of the magnetic fluctuations.  The coherent spectrum leads to a clear break because the structures cover a large number of frequencies up to their characteristic frequency, close to $f_b$. At $f_b$, their contribution to the spectrum drastically drops and the slope decrease sharply. 

The spectra and their slopes also provide indications on the presence of Alfv\'{e}n vortices. Indeed,  Alfv\'{e}n vortices have a specific  spectral shape because of their magnetic topology \citep{Alexandrova2008b}. The magnetic field of a monopolar vortex is located within a circle of radius $a$ \citep{Petviashvili1992}. This creates a discontinuity in the current density at the edge of the vortex, which implies a spectrum $f^{-2}$ for the current density. Therefore the magnetic field spectrum follows a power law of $f^{-4}$. So the $\sim f^{-4}$ spectrum at high frequencies ($f>f_b$) can be explained both by the current discontinuity spectrum of Alfv\'{e}n vortices and by the lack of contribution from other structures.

The overall spectrum is an average of the coherent and non-coherent areas, whose shape depends partly on the percentage of coherent events in the interval. 
We define the percentage of coherent events for a  given couple of magnetic field components $(B_i, B_j)$ as the
ratio between coherent and total times. It follows that the contribution of ion-cyclotron waves in the global spectrum is of the order of $20\%$, lower than the contribution of coherent structures (around $40\%$).

\begin{figure}
\begin{center}
\includegraphics[width=\simplesize]{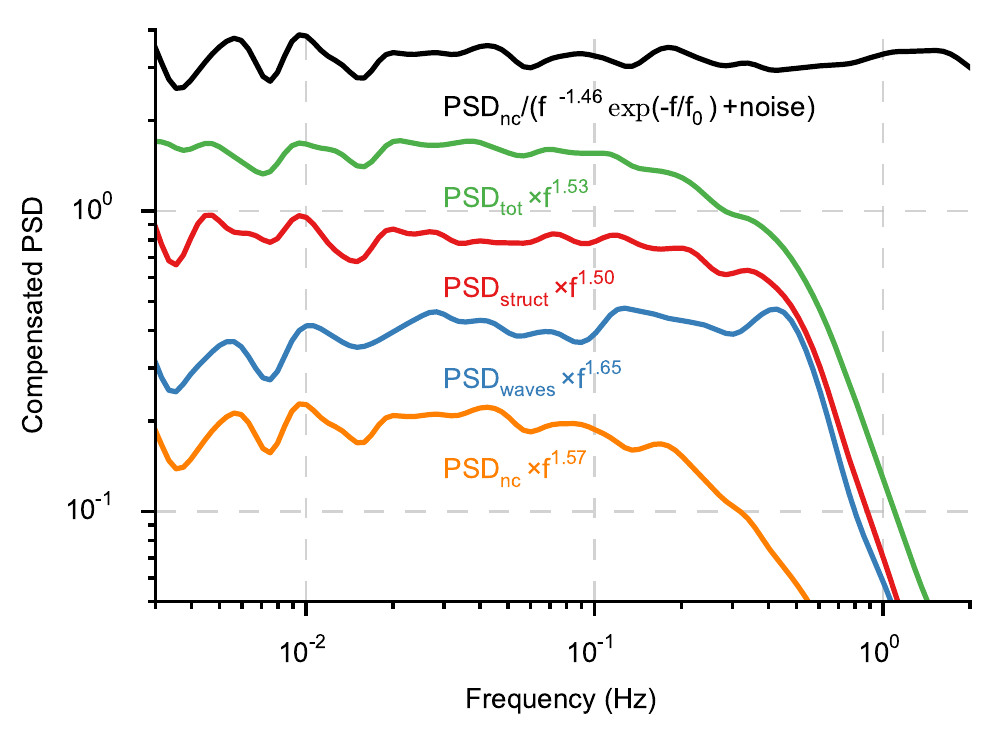}
\caption{Compensated spectra of Figure~\ref{fig9}: non-coherent spectrum compensated by the exponential fit (black) and each spectra compensate by their respective power law in the inertial range for comparison (same color code as in Figure~\ref{fig9}).}
\label{fig10}
\end{center}
\end{figure}

Figures~\ref{fig9}  and ~\ref{fig10} resume our findings regarding the turbulent spectrum around ion scales in the fast solar wind stream studied here:  It results from
the superposition of waves (solid blue), coherent structures (solid red) and non-coherent fluctuations (solid orange). 

The non-coherent and not polarized fluctuations have a spectrum without any break. It  can be modeled by a power law multiplied by an exponential cut-off $E(f)=E_0f^\alpha\exp(-f/f_0)$, with $E_0=6.4 \times 10^{-2}$ (considering frequencies in Hz and the magnetic field in nT), $\alpha=-1.46$ and $f_0=0.31$~Hz (see the dashed black line for the fit). This model  describes the evolution of the whole spectrum over 3 decades with a minimum number of free parameters.
It is better seen from the compensated spectrum of non-coherent fluctuations, shown in  Figure~\ref{fig10} by the black solid line (the other compensated spectra are shown here as well). One can see that it is flat  over all measured frequencies,  between $3 \times 10^{-3}$ and $2$~Hz, indicating that the model works quite well. 
  However the frequency range above the exponential cut-off frequency ($f>f_0$) is not large enough to conclude on the meaning of this adjustment and it will be necessary to consider both numerical simulations and observations over an extended frequency range to clearly understand phenomena at work and estimate if this exponential decay is due to any kind of dissipation.

\section{Summary and Conclusion}
\label{sec:conclusion}

In this paper we tried to understand which physical phenomena govern the transition between inertial and kinetic ranges and why this transition is highly variable. To address these questions, we selected a fast solar wind interval for which the spectrum exhibits a clear spectral break and a very steep slope of $-4$ at smaller scales (up to the noise level of the Wind/MFI instrument at 2-3~Hz). The choice of this interval was done in order to determine the conditions that give this specific spectral shape, not always observed in the solar wind.

Our analysis, based on phase coherency in time and scales calculated using the Morlet wavelet transform, shows the coexistence of (i) narrow-band incompressible left-handed circularly polarized waves with a central frequency at the spectral break frequency $f_b$; (ii)  coherent events in form of current sheets and Alfven vortex-like structures with characteristic scales close to $f_b$, but covering a wide range of frequencies $\sim[5\times 10^{-2},1]$~Hz starting one decade before $f_b$ in the inertial range; (iii)  non-coherent and non-polarized component of turbulence.

A detailed multi-instrumental analysis shows that the circularly polarized waves are observed when the flow to field angle $\theta_{BV}$ is close to zero. They propagate nearly parallel to the mean magnetic field ${\bf B_0}$ (i.e. $k_{\|}\gg k_{\perp}$) and their appearance is consistent with (1) favourable conditions for the development of the 
AIC instability (low ion beta $\beta_i\simeq 0,2$ and high ion temperature anisotropy $T_{\perp}/T_{\|}\simeq 4$) and (2) a low turbulence background level. The decrease of the background can be a consequence of the alignment between ${\bf B_0}$ and the solar wind velocity during wave events, see \citep{Horbury2008, Podesta2009}. The individual spectrum containing only these AIC waves exhibits a bump in the spectrum around $f_b$.

The coherent structures observed here (in particular current sheets and Alfv\'en vortex-like structures) are also mostly Alfv\'enic, i.e. with principal fluctuations perpendicular to ${\bf B}_0$. However, they have small compressible component with amplitude $\delta B_{\|}\ll \delta B_{\perp}$. Their geometry seems to be consistent with $k_{\perp}\gg k_{\|}$ anisotropy (however, to be sure, multi-satellite analysis should be done).   

The presence of current sheets and magnetic vortices in plasma turbulence is not new: numerous numerical simulations show this purpose. Current sheets are also largely observed in the solar wind turbulence, e.g. \citet{Greco2010, Perri2012}. 
However, regarding magnetic vortices,  there are only few examples of clear identification of such structures in space plasma turbulence. Alfv\'en vortices at ion spectral break scale have been identified in the Earth and Saturn magnetosheaths turbulence just behind a quasi-perpendicular portion of the bow-shocks \citep{Alexandrova2006, Alexandrova2008c}.
Kinetic Alfv\'en vortices (at scales smaller then the ion scales) have been observed in the Earth's cusp region by \citet{Sundkvist2005}. Signatures of large scales ($\sim 3$ minutes time scale, i.e. $f\ll f_b$) Alfv\'en vortices in the fast solar wind have been reported by \citet{Verkhoglyadova2003}. \citet{Roberts2013} showed an indirect indication of the presence of such vortices at ion scales: the authors applied k-filtering technique for a time interval of a fast wind stream and got dispersion relation which can be interpreted as oblique kinetic Alfv\'en waves (KAWs) or as convected structures, such as Alfv\'en vortices. Here, we show the presence of high amplitude localized magnetic fluctuations at ion scales which can be described by an Alfv\'en vortex model \citep{Petviashvili1992}. These observations may indicate that intermittency in the solar wind turbulence is not only related to planar structures (like current sheets) but there may also be filamentary structures like vortices.

As far as the observed vortex like structures are among the most energetic structures in the signal, one cannot neglect their influence to the observed total spectrum with a clear break and a $-4$ power-law at smaller scales. As was shown in \citet{Alexandrova2008b}, the monopole Alfv\'en vortex has a current discontinuity on its boundary and this gives a spectrum of the current  with a $-2$ power-law at scales smaller than the vortex radius. This corresponds to a $-4$ spectrum of magnetic field, as far as ${\bf j}= {\bf \nabla \times B} \sim k B$. At large scales, the vortex spectrum is flat and it has a knee at the scale of its radius. Indeed, as we show in the present study, the individual spectrum of the coherent structures exhibits a flattening  at $f<f_b$, a clear knee at $f_b$ and $-4$ power-law at $f>f_b$, consistent with the spectral properties of a monopole Alfv\'en vortex. Even if the coherent spectrum is similar to the one of the vortex, we must not neglect the role of other structures. Indeed, the sharp drop in the power spectrum can be seen not only as the contribution of the current discontinuity spectrum of  Alfv\'en vortices but also as the consequence of the lack of contribution from other coherent structures.

The third turbulence component, present in the analyzed time interval is not coherent nor polarized and it does not exhibit a spectral break at $f_b$ at all. Its individual spectrum has a smooth decreasing spectrum around ion scales well fitted by a power law multiplied by an exponential cut-off, $E(f)\propto f^\alpha\exp(-f/f_0)$ with $\alpha=-1.46$ and $f_0=0.31$~Hz. 

Consequently, ion scales can not be described by a single physical process and therefore, it is impossible to associate only one characteristic scale to the break frequency $f_b$. In particular,  this may explain as well, why different characteristic scales, such as $\rho_i$, $\lambda_i$ \citep{Chen2014}, and resonant wavenumber of AIC waves \citep{Bruno2014a} fit well $f_b$.

The presence of this break itself seems to be strongly related to a proportion and amplitude of  waves and coherent structures in the analyzed signal. The strong steepening with a $-4$ power-law seems to be simply the mix of spectrum of all the structures and maybe also an indication of the presence of Alfv\'en vortex monopoles. Therefore, following the results of \citet{Bruno2014b} and \citet{Smith2006}, the faster the solar wind (the stronger the energy transfer rate $\epsilon$), the steeper the spectrum at $f>f_b$, one may assume that in the fast wind there are more Alfv\'enic structures such as Alfv\'en vortices (or maybe simply more intense structures at ion scales) than in the slow wind, and in the presence of these structures the energy transfer rate is enhanced.

It will be interesting to analyze the role of the observed coherent structures in the problem of ion heating \citep{Smith2006, Matthaeus2008}, as already observed, for example, for much larger magnetic structures (hourly time scales), see for example \citep{Khabarova2015}. 

To summarize, variety of spectral shapes observed in the solar wind around ion scales can be explained by different number, intensity and duration of the coherent events, such as  waves and coherent structures,
which are themselves depending upon local plasma parameters (for waves) and non-local generation processes for coherent structures.

It is important to clarify the limitation of our study. As the coherency technique only detects phase correlated oscillation between magnetic field components, this method can miss events which appear only in one magnetic field component. 
We wish to emphasize as well that the present study is done only for 2 hours of data in the fast solar wind. To arrive to firm conclusions, a larger statistical study of the link between phase coherence of turbulent fluctuations and spectral shape should be done. This will be a subject of our future work.

\acknowledgments

The authors would like to thank Denise Perrone, Andr\'{e} Mangeney, Catherine Lacombe and Milan Maksimovic for useful discussions. The WIND data were obtained from the GSFC/SPDF CDAWeb interface at http://cdaweb.gsfc.nasa.gov/. The WIND MFI and SWE team are gratefully acknowledged for the magnetic field and proton data. This research made use of NASA's Astrophysics Data System, as well as matplotlib, a Python library for publication quality graphics \citep{Hunter:2007} and SciPy \citep{jones_scipy_2001}.

\appendix
\section{A. Continuous wavelet transform: spectrum and polarization}

\subsection{Spectrum and its confidence interval}

A continuous wavelet transform, e.g. \citep{Farge1992, Torrence1998}, allows to obtain from a time series $ B_j (t = t_i) $, with $ t_i = t_0 + i \delta t $ and $ i \in [| 0, N-1 |] $, the Power  Spectral Density (PSD) as a function of the frequency $ f $ in the spacecraft frame. This method has already been applied in previous studies, see e.g. \citet{Alexandrova2008a}. Using the Morlet wavelet transform of the $B_j(t)$ time series, we calculate the complex coefficient $W_j(f, t) $ as a function of the frequency in the spacecraft frame and the time $ t $. The total PSD, sum of the PSD of three components is then written as:
\begin{equation}\label{total_psd}
\begin{array}{ll}
E(f) &= \sum_{j=(x,y,z)}\frac{2\delta t}{N}\sum_{i=0}^{N-1} |W_j(f,t_i)|^2 \\
       &= \sum_{j=(x,y,z)}2\delta t \langle|W_j(f,t)|^2 \rangle_t.
\end{array}
\end{equation}

It is then necessary to determine if the wavelet spectrum is a good estimation of the true spectrum (and determine error bars for the wavelet spectrum). To do so,
let $\mathcal{W}^2(f, t)$ be the true wavelet power (the wavelet power in the ideal case) and $|W(f,t_i)|^2$ the estimated one  (our measurements). Formally, the probability that the estimated power should be close to the true power is:
\begin{equation}
P\left(\chi^2_{\nu, \alpha/2} < \nu \frac{|W(f,t)|^2}{\mathcal{W}^2(f,t)} < \chi^2_{\nu, 1-\alpha/2}\right) = 1 - \alpha.
\end{equation}
$\chi^2_{\nu, \alpha/2}$ represents the value at which the $\chi^2_\nu$ cumulative distribution function with $\nu$ degree of freedom equals $\alpha/2$ ($\nu=1$ for real coefficients, $\nu=2$ for complex coefficients) and where $\alpha$ is the desired significance ($\alpha=0.05$ for the $95\%$ confidence interval). The confidence interval for the wavelet power is then:  
\begin{equation}
\frac{\nu |W(f,t_i)|^2}{\chi^2_{\nu, 1-\alpha/2}} < \mathcal{W}^2(t,f) < \frac{\nu |W(f,t_i)|^2}{\chi^2_{\nu, \alpha/2}}.
\end{equation}
The wavelet spectrum is the average of more or less independent wavelet coefficients. Indeed, for the continuous wavelet transform, wavelet coefficients are correlated over a certain time which depends on the mother wavelet and which is characterized by the decorrelation length $\tau=\gamma s$ where $s$ is the scale ($s=\frac{1}{1.03 f}$ for the Morlet wavelet) and $\gamma$ is the decorrelation factor ($\gamma =2.32$ for the Morlet wavelet). This implies that if $\bar{\mathcal{W}}^2(f)$ is the true spectrum
 and $\bar{W}^2(f)=\frac{1}{N}\sum_{n=0}^{N-1} |W(f, t_n)|^2$ the estimated wavelet spectrum then the spectrum confidence interval is given by:
\begin{equation}
\label{confidence_limits}
\frac{\bar{\nu} \bar{W}^2(f)}{\chi^2_{\bar{\nu}, 1-\alpha/2}} < \bar{\mathcal{W}}^2(f) < \frac{\bar{\nu} \bar{W}^2(f)}{\chi^2_{\bar{\nu}, \alpha/2}}
\end{equation}
where $\bar{\nu}=\nu N \delta t / \tau$ is the new degree of freedom (which takes into account the number of coefficients used for averaging and the correlation between these coefficients) with $\delta t$ the time step.

For 1h of Wind/MFI measurements in the solar wind, it gives for one component : $0.86<\frac{\bar{\mathcal{W}}^2}{\bar{W}^2}<1.18$
at $f=0.1$~Hz and $0.95<\frac{\bar{\mathcal{W}}^2}{\bar{W}^2}<1.05$ for $f=1$~Hz. Note, that the confidence interval decreases with frequency.

\subsection{Polarization of magnetic fuctuations}

As indicated above, the Morlet wavelet coefficients $W_j (f, t) $ are complex. The argument $ \phi_j(f, t) = arg(W_j(f, t)) \; [2 \pi] $ of the complex coefficients can be interpreted as the local phase of the signal at a time $t$ and a frequency $f$ \citep{Grinsted2004}. The relative phasing between two time series, for example, $B_x(t)$ and $ B_y (t)$ is given by $\Delta \Phi_{xy} (f, t) = \phi_x (f, t) - \phi_y (f, t) $. Let $ \textbf{e}_x $, $ \textbf{e}_y $ and $ \textbf{e}_z $ be a direct trihedron, then the relative polarization with respect to the z-axis is given by
\begin{equation}
\label{eq:phase_diff}
\Delta \Phi_{xy}(f,t) =    \left\{
\begin{array}{ll}
\pi/2 &[2\pi] \rightarrow \mbox{right handed} \\
0 &\:[\pi] \rightarrow \mbox{linear} \\
-\pi/2 &[2\pi] \rightarrow \mbox{left handed}
\end{array}
\right. \begin{array}{ll}\end{array}
\end{equation}

\section{B. Wavelet transform coherence and phase filtering}
\label{sec:phase_filtering_appendix}

Two signals are said to be coherent if they maintain a fixed phase relationship.
We use the Wavelet Transform Coherence (WTC see \citet{Grinsted2004}) to separate coherent areas of the signal. The WTC highlights local phase lock behavior and provides a good indication of the local correlation between the time series in the time-frequency space. 

Let us consider two magnetic field components, $B_i (t)$ and $B_j(t)$, with $i,j=x,y,z$. The coherence coefficient  $R_{ij}(f,t)$, which characterize phase coupling between $B_i (t)$ and $B_j(t)$ is defined using the Continuous Wavelet Transform 
of two signals:
\begin{equation}
\label{eq:coherence_map}
R^2_{ij}(f,t)=\frac{|S(fW_{i}(f,t)W_{j}^*(f,t))|^2}{S(f|W_i(f,t)|^2)\cdot S(f|W_j(f,t)|^2)}
\end{equation}
where $W_{i(j)}(f,t)$ are  complex wavelet coefficients of $B_{i(j)}(t)$  \citep{Farge1992}, $S$ is a smoothing operator defined by $S(W(f,t))=S_{freq}( S_{time}(W(f,t)) )$ with $S_{freq}(W(f,t))=W(f,t)c_1^{-t^2f^2/2}$ the smoothing operator over frequencies, and $S_{time}(W(f,t))=W(f,t)c_2\Pi(0.6/f)$ over time. $c_1$ and $c_2$  are numerically determined normalization constants (see \citet{Grinsted2004}) and $\Pi$ the rectangular function. The factor of $0.6$ is the empirically determined scale decorrelation length for the Morlet wavelet \citep{Torrence1998}. The normalization through the local average operator $S$ allows to consider not only the high-amplitude events with coupled phases, but all the phase coupled events.
  By definition $R_{ij}(f,t)$ is between 0 (no coherence) and 1 (full coherence).

To remove the maximum of fortuitous coherence between two component $(B_i,B_j)$ we need to determine the statistical significance level of the WTC. For this purpose we construct $N$ $(=100)$  surrogate time series for each component from the original data for which we randomize the phases as done in \citep{Hada2003,Koga2003}. So, the significance threshold is defined as

\begin{equation}
\label{eq:coherence_threshold}
R_{ij}^{threshold}=\overline{\mathstrut R}_{ij}^{random}+2\sigma_{ij}^{random},
\end{equation}
using the mean
\begin{equation}
\overline{\mathstrut R}_{ij}^{random}(f,t)=\frac{1}{N}\sum_{k=1}^{N} R_{ij}^{(k)}(f,t), \; N=100
\end{equation}
and the standard deviation of the WTC for the 100 surrogate datasets
\begin{equation}
\sigma_{ij}^{random}(f,t)=\sqrt{ \frac{1}{N}\sum_{k=1}^{N} (R_{ij}^{(k)}(f,t)-\overline{\mathstrut R}_{ij}^{random}(f,t))^2}.
\end{equation}
If $R_{ij}^{threshold}>1$, we take $R_{ij}^{threshold}=1$ because $R$ can not be greater than 1. 

We call coherent times $T^c_{ij}$ the ensemble of time points, such that two magnetic field components, $B_i (t)$ and $B_j(t)$, are coupled in phase. These times are
defined as the time set $t$ which verifies
\begin{equation}
R_{ij}(f_b,t) \ge R^{threshold}_{ij}(f_b,t).
\end{equation}
The complementary set, $T^{nc}_{ij}$, corresponds to the non-coherent times. 

The individual spectra of coherent  ($c$) and non coherent ($nc$) parts of the signal are then defined as
\begin{equation}
\label{eq:individual_spectrum}
E^{c/nc}_{ij}(f)=\sum_{k=(x,y,z)}2\delta t \langle|W_k(f,t)|^2\rangle_t \mbox{ with } t \in T^{c/nc}_{ij}\mbox{.}
\end{equation}

\bibliography{biblio.bib}

\end{document}